# Quantum theory of preparation and measurement

DAVID T. PEGG

School of Science, Griffith University, Nathan, Brisbane,

Q 4111, Australia

STEPHEN M. BARNETT and JOHN JEFFERS

Department of Physics and Applied Physics, University

of Strathclyde, Glasgow G4 0NG, Scotland

---

Author for correspondence: D. T. Pegg

Phone: +61 7 3875 7152, Fax: +61 3875 7656, Email: D.Pegg@sct.gu.edu.au

**Abstract.** The conventional postulate for the probabilistic interpretation of quantum mechanics is asymmetric in preparation and measurement, making retrodiction reliant on inference by use of Bayes' theorem. Here we present a more fundamental symmetric postulate from which both predictive and retrodictive probabilities emerge immediately, even where measurement devices more general than those usually considered are involved. We show that the new postulate is perfectly consistent with the conventional postulate.

## 1. Introduction

The conventional formalism of quantum mechanics based on the Copenhagen interpretation is essentially predictive. We assign a state to a system based on our knowledge of a preparation event and use this state to predict the probabilities of outcomes of future measurements that might be made on the system. If we have sufficient knowledge to assign a pure state, then this state contains the maximum amount of information that nature allows us for prediction. With less knowledge, we can only assign a mixed state. This formalism works successfully. Sometimes, however, we may have knowledge of the result of a measurement and wish to retrodict the state prepared. A particular example of this is in quantum communication where the recipient receives a quantum system that the sender has prepared and sent. If the prepared state has not evolved at the time of measurement to an eigenstate of the operator representing the recipient's measurement, then the best retrodiction that the recipient can make is to calculate probabilities that various states were prepared. While it is possible to do this by using the usual predictive formalism and inference based on Bayes' theorem [1], this is often quite complicated. Aharonov *et al.* [2], in investigating the origin of the arrow of time, formulated a retrodictive formalism that involves assigning a state based on knowledge of the measurement outcome. This state is assigned to the system just prior to the measurement and evolves backward in time to the preparation event. While this formalism seems to offer a more direct means of retrodiction, Belinfante [3] has argued that the formalism is only valid in very particular circumstances that essentially involve the prepared states, which in his case are eigenstates of a preparation operator, having a flat *a priori* probability distribution. While the lack of preparation knowledge associated with such an unbiased distribution is sometimes applicable, in general it is not.

In our recent work [4 - 6] we have found quantum retrodiction useful for a variety of applications in quantum optics. Furthermore the formalism can be generalised so as to be applicable when there is not a flat *a priori* probability distribution for the prepared states by using Bayes' theorem [6]. The price of this generalisation appears to be a loss in symmetry between preparation and measurement. In this paper we adopt a formal approach to investigate this question more closely. We find that we can replace the usual measurement postulate of the probability interpretation of quantum mechanics by a fundamental postulate that is symmetric in measurement and preparation. This allows us to formulate a more general theory of preparation and measurement than that of the conventional formalism and makes clear the relationship between the predictive and retrodictive approaches. The new postulate also allows us to see clearly Belinfante's argument in an appropriate perspective. We show that our new postulate is entirely in accord with the conventional postulate. The retrodictive formalism results in the same calculated experimental outcomes of quantum mechanics as does the conventional approach despite the fact that we ascribe a different state to the system between preparation and measurement.

## 2. Preparation and measurement devices

We consider a situation where Alice operates a device that prepares a quantum system and Bob does subsequent measurements on the system and records the results. The preparation device has a readout mechanism that indicates the state the system is prepared in. We associate a preparation readout event *i*, where $i = 1, 2, \cdots$, of the preparation device with an operator $\hat{\Lambda}_i$ acting on the state space of the system, which we call a preparation device operator (PDO). This operator not only represents the prepared state but also contains information about any bias in its preparation. A bias might arise, for example, because the device may not be able to produce certain states or Alice may choose rarely to prepare other states. We describe the operation of the preparation device mathematically by a set of PDOs. The measurement device also has a readout mechanism that shows the result of the measurement. We associate a measurement readout event $j$, where $j = 1, 2, \cdots$, of the measurement device with a measurement device operator (MDO) $\hat{\Gamma}_j$ acting on the state space of the system. This operator represents the state of the system associated with the measurement and contains information about any bias on the part of Bob or the device in having the measurement recorded. For example for a von Neumann measurement the MDO would be proportional to a pure state projector. We describe the operation of the measurement device mathematically by a set of MDOs. In general the operators $\hat{\Lambda}_i$ need not be orthogonal to each other, and nor do the operators $\hat{\Gamma}_j$.

In order to eliminate the complication of time evolution we assume for now that the system does not change between preparation and measurement. For example, there may not be a sufficiently long time between preparation and measurement for evolution to occur. In an experiment Alice chooses a state to prepare and, when the readout mechanism indicates that this state has been successfully prepared, the preparation readout event *i* is automatically sent to a computer for recording. Bob then measures the system. If he chooses, he may then send the measurement readout event *j* obtained to the computer for recording. If the computer receives a record from both Alice and Bob it registers combined event $(i, j)$. The measurement device may not produce a readout event corresponding to every possible preparation event and different preparation events may lead to the same measurement readout event. There is not necessarily a uniform probability that Bob will record all readout events. The preparation device may be capable of preparing only a limited number of states. There is not necessarily a uniform probability that Alice will choose to prepare all these states. The experiment is repeated many times with Alice choosing states to prepare as she wishes and Bob recording the measurement readout events he chooses. The computer produces a list of combined events $(i, j)$ from to each experiment, from which various occurrence frequencies can be found.

We may wish to *predict* the measurement result that will be recorded in a particular experiment on the basis of our knowledge of the actual preparation event $i$ and our knowledge of the operation of the measuring device, that is, of the set of MDOs. Because of the nature of quantum mechanics, we usually cannot do this with certainty, the best we can do is to calculate the probabilities that various possible states will be detected and recorded by Bob. Similarly the best we can do in *retrodicting* the preparation event recorded by Alice in a particular experiment on the basis of our knowledge of the recorded measurement event $j$ and our knowledge of the set of PDOs for the preparation device, is to calculate probabilities for possible preparation events. Our aim in this paper is to postulate a fundamental relationship that allows us to calculate such predictive and retrodictive probabilities, which could then be compared with the occurrence frequencies obtained from the collection of combined events $(i, j)$ recorded by the computer.  In this way a theory of quantum retrodiction is verifiable experimentally.

Difficulties have arisen in studying retrodiction [3] because the usual formulation of quantum mechanics is predictive.  That is, measurement theory is formulated in terms of predicting measurement outcomes.  In order to keep preparation and measurement as well as prediction and retrodiction on a symmetric footing, it is convenient to reformulate the probability interpretation of quantum mechanics by means of postulate (1) below. We show that this leads to the conventional asymmetric predictive postulate and, as an assurance that our approach is perfectly equivalent to predictive theory, in the Appendix we derive postulate (1) from conventional measurement theory.

## 3.  Fundamental postulate

A sample space of mutually exclusive outcomes can be constructed from the collection of recorded combined events by identifying these events with points of the space so that identical events are identified with the same point.  A probability measure assigns probabilities between zero and one to the points such that these probabilities sum to unity for the whole space. The probability assigned to a point $(i, j)$ is proportional to the number of combined events $(i, j)$ identified with that point, that is, to the occurrence frequency of the event $(i, j)$.  Our fundamental *postulate* in this paper for the probabilistic interpretation of quantum mechanics is that the probability associated with a particular point $(i, j)$ in this sample space is

$$P^{\Lambda\Gamma}(i, j) = \frac{\mathrm{Tr}(\hat{\Lambda}_i \hat{\Gamma}_j)}{\mathrm{Tr}(\hat{\Lambda}\hat{\Gamma})} \qquad (1)$$

where the trace is over the state space of the system and

$$\hat{\Lambda} = \sum_i \hat{\Lambda}_i \tag{2}$$

$$\hat{\Gamma} = \sum_j \hat{\Gamma}_j \quad . \tag{3}$$

In order to ensure that no probabilities are negative, we assume that $\hat{\Lambda}_i$ and $\hat{\Gamma}_j$ are non-negative definite. If a combined event from an experiment chosen at random is recorded then expression (1) is the probability for that event to be (*i*, *j*). That is, expression (1) is the probability that the state prepared by Alice corresponds to $\hat{\Lambda}_i$ and the state detected by Bob corresponds to $\hat{\Gamma}_j$, given that Bob has recorded the associated measurement event. The essence of the postulate lies in the numerator of (1); the denominator simply ensures that the total probability for all the recorded mutually exclusive outcomes is unity. We note that the fundamental expression (1) only requires $\hat{\Lambda}_i$ and $\hat{\Gamma}_j$ to be specified up to an arbitrary constant. That is, we can multiply all the $\hat{\Gamma}_j$ by the same constant without affecting $P^{\Lambda\Gamma}(i, j)$ and similarly for $\hat{\Lambda}_i$. We use this flexibility later to choose $\hat{\Gamma}_j$ for convenience such that $\hat{1} - \hat{\Gamma}$ is non-negative definite, where $\hat{1}$ is the unit operator. We shall also use this flexibility in choosing $\hat{\Lambda}_i$.

From (1) we can deduce the following probabilities:

$$P^{\Lambda\Gamma}(i) = \sum_j P^{\Lambda\Gamma}(i,j) = \frac{\mathrm{Tr}(\hat{\Lambda}_i\hat{\Gamma})}{\mathrm{Tr}(\hat{\Lambda}\hat{\Gamma})} \tag{4}$$

$$P^{\Lambda\Gamma}(j) = \frac{\mathrm{Tr}(\hat{\Lambda}\hat{\Gamma}_j)}{\mathrm{Tr}(\hat{\Lambda}\hat{\Gamma})} \tag{5}$$

$$P^{\Lambda\Gamma}(j \mid i) = \frac{P^{\Lambda\Gamma}(i,j)}{P^{\Lambda\Gamma}(i)} = \frac{\mathrm{Tr}(\hat{\Lambda}_i\hat{\Gamma}_j)}{\mathrm{Tr}(\hat{\Lambda}_i\hat{\Gamma})} \tag{6}$$

$$P^{\Lambda\Gamma}(i \mid j) = \frac{\mathrm{Tr}(\hat{\Lambda}_i\hat{\Gamma}_j)}{\mathrm{Tr}(\hat{\Lambda}\hat{\Gamma}_j)} \tag{7}$$

Expression (4) is the probability that, if an experiment chosen at random has a recorded combined event, this event includes preparation event *i*. Likewise (5) is the probability that the recorded combined event includes the measurement event *j*. Expression (6) is the probability that, if the recorded combined event includes event *i*, it also includes event *j*. That is, it is the probability that the event recorded by Bob is the detection of the state corresponding to $\Gamma_j$ if the state prepared by Alice in the experiment corresponds to $\hat{\Lambda}_i$. Expression (6) can be obtained by limiting the sample space to those events containing *i* and is essentially Bayes' formula [7]. Likewise (7) is the probability that the state prepared by Alice corresponds to $\hat{\Lambda}_i$ if the event recorded by Bob is the detection of the state corresponding to $\Gamma_j$.

Expression (6) can be used for prediction. In order to calculate the required probability from our knowledge of the PDO $\hat{\Lambda}_i$ associated with the preparation event *i* we must also know every possible MDO $\hat{\Gamma}_j$ that is, we must know the mathematical description of the operation of the measuring device. Similarly we can use (7) for retrodiction if we know $\hat{\Gamma}_j$ and all the $\hat{\Lambda}_i$ of the preparation device.

## 3. Unbiased devices

### 3.1. *A priori probability*

Of all the states that Alice might prepare, there is an *a priori* probability, which is independent of the subsequent measurement, that she chooses a particular one. For $P^{\Lambda\Gamma}(i)$ in (4) to represent this *a priori* probability the expression for $P^{\Lambda\Gamma}(i)$ must be *independent of the operation of measurement device*. A specific condition must be imposed on the measuring device and its operation to do this. This condition is that the set of MDOs describing the operation of the measurement device must be such that their sum $\hat{\Gamma}$ is proportional to the identity operator on the state space of the system, that is

$$\hat{\Gamma} = \gamma \hat{1} \tag{8}$$

say where $\gamma$ is a positive number. Then we can replace $\hat{\Gamma}$ in the numerator and denominator in (4) by the unit operator and the influence of $\hat{\Gamma}$ is removed from the expression, making $P^{\Lambda\Gamma}(i)$ equal to $P^{\Lambda}(i)$ where the latter is defined as

$$P^{\Lambda}(i) = \frac{\mathrm{Tr}\hat{\Lambda}_i}{\mathrm{Tr}\hat{\Lambda}} \tag{9}$$

Expression (9) is the $\Gamma_j$-independent, *a priori*, probability that the state prepared by Alice corresponds to $\hat{\Lambda}_i$.

It is useful also to define an operator

$$\hat{\rho}_i = \frac{\hat{\Lambda}_i}{\mathrm{Tr}\hat{\Lambda}_i} \quad . \tag{10}$$

The trace of $\hat{\rho}_i$ is unity so these non-negative operators are *density operators* describing the states Alice may prepare. From the definitions (9) and (10) we can write $\hat{\Lambda}_i$ as proportional to $P^{\Lambda}(i)\hat{\rho}_i$. The constant of proportionality always cancels in the expressions for the various probabilities so there is no loss of generality in taking this constant to be unity. Then we have

$$\hat{\Lambda}_i = P^{\Lambda}(i)\hat{\rho}_i \quad . \tag{11}$$

We see explicitly from (11) how the PDO $\hat{\Lambda}_i$, as well as representing the prepared state, also contains information about the bias in its preparation. The biasing factor is simply the *a priori* preparation probability.

From (9), (11) and (2) we see that $\hat{\Lambda}$ has unit trace so it also is a density operator given by

$$\hat{\Lambda} = \hat{\rho} = \sum_i P^{\Lambda}(i)\hat{\rho}_i \quad . \tag{12}$$

This is the best description we can give of the state prepared by Alice if we do not know which particular preparation or measurement event took place but we do know the possible states she can prepare and the *a priori* probabilities associated with each.

3.2. *Unbiased measurements*

We call the operation of a measurement device for which (8) is true, and thus $P^{\Lambda\Gamma}(i) = P^{\Lambda}(i)$, *unbiased*. Not all measurements are unbiased, as we shall discuss later, but for now we shall focus on measuring devices with unbiased operations. For these it is convenient to define

$$\hat{\Pi}_j = \frac{\hat{\Gamma}_j}{\gamma} \quad . \tag{13}$$

From (6), (8) and (10) we then obtain

$$P^{\Lambda\Gamma}(j\,|\,i) = \mathrm{Tr}(\hat{\rho}_i \hat{\Pi}_j) \tag{14}$$

From (13) and (8) the sum of $\hat{\Pi}_j$ is the unit operator, so these non-negative operators form the elements of a *probability operator measure* (POM) [8]. Our result (14) is the *fundamental postulate of quantum detection theory* [8]. Thus our postulate (1) reduces to the conventional postulate for unbiased measurements. Expressions (14) and

(10) allow us to identify the PDO $\hat{\Lambda}_i$ for the preparation of a pure state as being proportional to the corresponding pure state projector.

It is worth remarking on the asymmetry of (14) in that the PDO has become a density operator and the MDO has become a POM element. In the simple case where both the PDO and the MDO are pure state projectors, as for a von Neumann measurement of a pure state, symmetry is restored. In general, however, density operators and POM elements have quite different normalisation properties. The asymmetry in preparation and measurement, and hence a time asymmetry, does not arise here through some basic asymmetry in quantum mechanics. Rather it arises from our request that the probability for Alice's choice of preparation event be independent of subsequent measurement. This is usually an implicit assumption in the conventional, that is predictive, probability interpretation of quantum mechanics. The apparent asymmetry is reinforced by adopting (14) as a fundamental postulate of measurement theory as done for example by Helstrom [8].

A simple, but important, example of unbiased measurement is the case where no measurement is made. For example the measuring device might not interact with the system at all and thus gives a meter reading of zero for all prepared states. As there is only one measurement readout event, there is only one MDO $\hat{\Gamma}_j = \hat{\Gamma}$. The only probability that we can assign to a preparation event if we do not know the preparation readout event and if we have made no measurement on the system is the *a priori* probability $P^{\Lambda}(i)$. Thus if we calculate the retrodictive probability $P^{\Lambda\Gamma}(i\,|\,j)$ on the basis of the no-measurement state, then we must obtain $P^{\Lambda}(i)$. From (7) and (9), $\hat{\Gamma}$ must therefore be proportional to the unit operator and so the measurement must be unbiased.

The single POM element for the measuring device must be $\hat{1}$ to ensure that the sum of the elements is the unit operator.

The operation of most ideal measuring devices is usually unbiased, but this is not always the case. In [6] we discussed two-photon interference for photons from a parametric down-converter where results from higher-number states are discarded. Another example is in the operational phase measurements of Noh *et al.* [9]. Here certain photo-detector readings are not recorded because they do not lead to meaningful values of the operators being measured. The probabilities used for the experimental statistics are then suitably renormalised.

### 3.3. *Unbiased preparation*

We say in general that the operation of a preparation device is unbiased if the PDOs $\hat{\Lambda}_i$ are proportional to $\hat{\Xi}_i$ where

$$\sum_i \hat{\Xi}_i = \hat{1} \tag{15}$$

that is, if the operators $\hat{\Xi}_i$ form the elements of a preparation device POM. Then, for a preparation device with an unbiased operation, $P^{\Lambda\Gamma}(j)$ is independent of $\hat{\Lambda}_i$ and

$$P^{\Lambda\Gamma}(i \mid j) = \mathrm{Tr}(\hat{\Xi}_i \hat{\rho}_j^{\mathrm{retr}}) \tag{16}$$

where

$$\hat{\rho}_j^{\text{retr}} = \hat{\Gamma}_j / \text{Tr}\hat{\Gamma}_j. \tag{17}$$

A specific example of a preparation device with an unbiased operation is where Alice prepares a spin-half particle in the up or down state, each with a probability of one-half. The two preparation device operators $\hat{\Lambda}_{up}$ and $\hat{\Lambda}_{down}$ can then be taken as proportional to density operators given by the respective projectors $|up\rangle\langle up|$ and $|down\rangle\langle down|$. Then $\hat{\Lambda}$ is proportional to the unit operator on the state space of the particle and we find from (7) that

$$P^{\Lambda\Gamma}(up \mid j) = \text{Tr}(|up\rangle\langle up|\hat{\rho}_j^{\text{retr}}) \tag{18}$$

which gives the retrodictive probability that the state in which Alice prepared the particle was the up state if Bob detected the state $\hat{\rho}_j^{\text{retr}} = \hat{\Gamma}_j / \text{Tr}\hat{\Gamma}_j$. This is consistent with (16) with $\hat{\Xi}_{up} = |up\rangle\langle up|$.

Many preparation devices have biased operations, so (16) is not applicable to them. For example the preparation of a field in a photon number state may be constrained through limited available energy. In this case the set of PDOs would not include projectors for higher photon number states and thus could not sum to be proportional to the unit operator in the whole state space of the field. Alternatively, Alice might prepare the spin-half particle in the up state or in an equal superposition of the up and down states only. For such situation we must use the more general form of the retrodictive probability (7).

## 4. Time evolution

In the conventional approach, when the state of system changes unitarily between preparation and measurement, we replace $\hat{\rho}_i$ by $\hat{\rho}_i(t_m) = \hat{U}\hat{\rho}_i\hat{U}^\dagger$ in the appropriate probability formulae where $\hat{U}$ is the time evolution operator between the preparation time $t_p$ and the measurement time $t_m$. Thus in this paper we replace $\hat{\Lambda}_i$ by $\hat{\Lambda}_i(t_m) = \hat{U}\hat{\Lambda}_i\hat{U}^\dagger$ while noting that $\mathrm{Tr}(\hat{U}\hat{\Lambda}_i\hat{U}^\dagger) = \mathrm{Tr}\hat{\Lambda}_i$. This is clearly consistent with (10) and yields the usual predictive formula (14) with $\hat{\rho}_i$ replaced by $\hat{\rho}_i(t_m)$. For the retrodictive probability replacing (7) we obtain, using the definition (17),

$$P^{\Lambda\Gamma}(i\,|\,j) = \frac{\mathrm{Tr}(\hat{U}\hat{\Lambda}_i\hat{U}^\dagger\hat{\rho}_j^{\mathrm{retr}})}{\mathrm{Tr}(\hat{U}\hat{\Lambda}\hat{U}^\dagger\hat{\rho}_j^{\mathrm{retr}})} \quad . \tag{19}$$

From the cyclic property of the trace we can rewrite this as

$$P^{\Lambda\Gamma}(i\,|\,j) = \frac{\mathrm{Tr}[\hat{\Lambda}_i\hat{\rho}_j^{\mathrm{retr}}(t_p)]}{\mathrm{Tr}[\hat{\Lambda}\hat{\rho}_j^{\mathrm{retr}}(t_p)]} \tag{20}$$

where $\hat{\rho}_j^{\mathrm{retr}}(t_p) = \hat{U}^\dagger\hat{\rho}_j^{\mathrm{retr}}\hat{U}$ is the retrodictive density operator evolved backwards in time to the preparation time. This is the retrodictive formula we obtained previously [6] using the conventional approach and Bayes' theorem [1]. We note that (20) can be interpreted as the state collapse taking place at the preparation time $t_p$. This arbitrariness in when we choose to say the collapse occurs is not confined to retrodiction. Even the conventional predictive formula obtained from (14) by replacing $\hat{\rho}_i(t_m)$ by $\hat{\rho}_i(t_m) = \hat{U}\hat{\rho}_i\hat{U}^\dagger$ can be rewritten as $\mathrm{Tr}(\hat{\rho}_i\hat{U}^\dagger\hat{\Pi}_j\hat{U})$ where $\hat{U}^\dagger\hat{\Pi}_j\hat{U}$ can be interpreted as an element of a POM describing the operation of a different measuring device for which the measurement event takes place immediately after the preparation time $t_p$.

## 5. Example

As an important example of our approach, we apply it in this section to the experimental situation envisaged by Belinfante [3]. After studying the work of Aharonov *et al.*[2], Belinfante came to the conclusion that retrodiction is only valid in very special circumstances. He examined the situation where a measurement device B makes von Neumann measurements with outcomes corresponding to a complete set of pure states $|b_j\rangle$. His preparation device, which prepares pure states $|a_i\rangle$, comprises a measuring device A making von Neumann measurements on a system in a state given by a density operator $\hat{\rho}_g$. The predictive probability that the state measured is $|b_j\rangle$ if the state prepared is $|a_i\rangle$ is $|\langle a_i|b_j\rangle|^2$. Belinfante argued that quantum theory would be time-symmetric in its probability rules if the retrodictive probability that the state prepared is $|a_i\rangle$, if the state measured is $|b_j\rangle$, is taken as $|\langle b_j|a_i\rangle|^2$, which is the retrodictive inverse of $|\langle a_i|b_j\rangle|^2$. These two expressions are equal. Belinfante concluded that retrodiction is

valid only if the mixed state of the system before measurement by A is uniformly "garbled", that is if the density operator $\hat{\rho}_g$ is proportional to the unit operator.

Let us examine this situation in terms of our formalism. The operation of the von Neumann measuring device B is unbiased so we can describe it by a set of PDOs which form a POM with elements

$$\hat{\Gamma}_j = \hat{\Pi}_j^b = |b_j\rangle\langle b_j|. \tag{21}$$

Similarly the operation of the measuring device A is described by the POM with elements $\hat{\Pi}_i^a = |a_i\rangle\langle a_i|$. The *a priori* probability for state $\hat{\rho}_i = |a_i\rangle\langle a_i|$ to be prepared is $\mathrm{Tr}(\hat{\rho}_g \hat{\Pi}_i^a)$. Thus from (11) we have

$$\hat{\Lambda}_i = \mathrm{Tr}\left(\hat{\rho}_g |a_i\rangle\langle a_i|\right)|a_i\rangle\langle a_i| \tag{22}$$

From (14), the predictive probability for an unbiased measuring device, we find that the probability that the state measured is $|b_j\rangle$ if the state prepared is $|a_i\rangle$ is $|\langle a_i|b_j\rangle|^2$. This agrees with Belinfante's result. However, the retrodictive probability (7) becomes, from (21) and (22)

$$P^{\Lambda\Gamma}(i\,|\,j) = \frac{\mathrm{Tr}\left(\hat{\rho}_g |a_i\rangle\langle a_i|\right)|\langle a_i|b_j\rangle|^2}{\sum_i \left[\mathrm{Tr}\left(\hat{\rho}_g |a_i\rangle\langle a_i|\right)|\langle a_i|b_j\rangle|^2\right]} \tag{23}$$

for the probability that the state prepared is $|a_j\rangle$ if the state measured is $|b_j\rangle$. This agrees with the result of Belinfante if, and only if, $\hat{\rho}_g$ is proportional to the unit operator.

From the above, we see that the difficulty with retrodiction raised by Belinfante is due to use of the retrodictive inverse of an inappropriate predictive formula. Belinfante effectively found $P^{\Lambda\Gamma}(i\,|\,j)$ by taking the retrodictive inverse of $P^{\Lambda\Gamma}(j\,|\,i)$ in (14). However (14) is valid only for unbiased measuring devices and its retrodictive inverse, which is given by (16), is only valid for unbiased preparation devices. It is not surprising then that Belinfante found his retrodictive formula only worked if $\hat{\rho}_g$ is proportional to the unit operator as this is precisely the condition needed to ensure that the PDOs (22) describe the operation of an unbiased preparation device. For biased preparation we must use the retrodictive inverse of the *more general* predictive formula (6) which is just (7) as used above. We conclude that retrodiction is valid for a general preparation device provided the correct formula is used.

## 6. Conclusion

Overall, the approach adopted in this paper to the probability interpretation of quantum mechanics puts preparation and measurement on a more equal footing than in the conventional approach where preparation is usually ignored and the measuring device is assumed to be unbiased. We have formulated our approach in terms of more general sets of non-negative definite operators than POMs. We have found that for an unbiased measuring device, for which the measuring device operators reduce to the elements of a POM, the preparation device operators can be written as density operators, absorbing the normalisation denominator in the general expression (6). This reduces (6) to (14), the conventional asymmetric postulate of quantum detection theory. Just as (14) is only applicable for unbiased measuring devices, its retrodictive inverse (16) is only applicable for unbiased preparation devices. These latter devices are unusual in practice, which leads to Belinfante's objection to retrodiction. A useful theory of retrodiction requires that allowance be made for bias in the preparation device. A fully symmetric probability interpretation of quantum mechanics would then also require allowance to be made for a biased measurement device as we have done in this paper.

As mentioned in the introduction, the retrodictive formalism results in the same calculated experimental outcomes of quantum mechanics as does the conventional approach based on the Copenhagen interpretation, despite the fact that we ascribe a different state to the system between preparation and measurement. In the conventional approach, the state assigned to the system contains the information needed to predict the outcomes of possible measurements on the system. In this sense, the conventional approach is essentially predictive in nature and is thus a legitimate part of the broader picture that also includes retrodiction. Indeed the conventional approach is sufficient in the sense that one can perform retrodictive probability calculations by using it together with Bayes' theorem. On the other hand, this approach is not necessary in that one could perform predictive probability calculations, albeit complicated, using the retrodictive formalism plus Bayes' theorem. Thus both the conventional and retrodictive formalisms should be viewed merely as means for calculating probabilities with one being more convenient than the other depending on the situation. We should also mention, however, that retrodiction also raises interesting philosophical questions if one wishes to ascribe a physical existence or reality to the state in the ontological sense. These issues go beyond trying to decide if the state of the system is "really" the predictive or the retrodictive state. In [5] it is shown that it is possible for the retrodictive state to be entangled for some situations where there is no entanglement in the predictive picture. In the predictive formalism, the Many-Worlds interpretation [10] depicts an increasing number of branching universes that include the different possible results of measurements as we go forward in time. In the retrodictive formalism a Many-Worlds interpretation should look very different. Presumably the branching will occur as we go backwards in time from the measurement to the preparation. We do not intend to pursue such questions here. As long as the retrodictive formalism yields the correct quantum mechanical probabilities, we view it as an acceptable and sometimes more convenient approach to quantum mechanics and shall leave the philosophical issues to metaphysics.

**Acknowledgments**

This work was supported by the Australian Research Council and the U.K. Engineering and Physical sciences Research Council. SMB thanks the Royal Society of Edinburgh and the Scottish Office Education and Lifelong Learning Department for the award of a Support Research Fellowship.

**Appendix**

In this appendix we derive our general postulate (1) from the standard predictive postulate (14). As we have already shown how (14) follows from (1), this establishes that (1) is both necessary and sufficient for the accepted probability interpretation of quantum mechanics.

The operation of the measuring device *M* used by Bob is described by the set of MDOs $\hat{\Gamma}_j$ with $j = 1, 2, \cdots$. As discussed earlier, we choose for convenience the arbitrary constant in $\hat{\Gamma}_j$ such that $\hat{1} - \hat{\Gamma}$ is non-negative definite. This allows us to define a set of non-negative definite operators $\hat{\Pi}_k$ by

$$\hat{\Pi}_j = \hat{\Gamma}_j \quad \text{for} \quad j = 1, 2, \cdots \qquad \text{(A 1)}$$

$$\hat{\Pi}_0 = \hat{1} - \hat{\Gamma} . \qquad \text{(A 2)}$$

It is clear from (3) that the operators $\hat{\Pi}_k$ sum to the unit operator and thus form the elements of a POM. We can use this POM to define the operation of another measuring device $\overline{M}$ which has precisely the same operation as that of *M*, except that it allows an extra measurement event $k = 0$ to be recorded. The readout for this event can be interpreted as "none of the events *j*". We can use the usual postulate corresponding to (14) to obtain the probability that measurement event *k* will be recorded by $\overline{M}$ if the system is prepared in state $\hat{\rho}_i$ as

$$P^{\Lambda\Pi}(k \mid i) = \mathrm{Tr}(\hat{\rho}_i \hat{\Pi}_k) \ . \qquad \text{(A 3)}$$

Thus

$$P^{\Lambda\Pi}(i,k) = \mathrm{Tr}(\hat{\rho}_i \hat{\Pi}_k) P^{\Lambda}(i) \qquad \text{(A 4)}$$

If Bob had used $\overline{M}$ in place of *M*, a sample space of combined events (*i, k*) would have been obtained that is larger than that of events (*i, j*) obtained with *M* in that it includes some extra points (*i*, 0). If these extra events are ignored, then the difference between the operations of $\overline{M}$ and *M* vanishes, so the restricted sample space of events $(i,k)$ with $k \neq 0$ will be the same as the sample space of events (*i, j*) for *M*. The probability $P^{\Lambda\Gamma}(i,j)$ will thus be equal to the probability of finding the event (*i*, *k*), with *k* not zero, in this restricted sample space. This probability will be equal to $P^{\Lambda\Pi}(i,j)$ with a normalisation factor to ensure that the total probability for the restricted sample space is unity. From (A4), (A1) and from the definition (3) we then have

$$P^{\Lambda\Gamma}(i,j) = \frac{\mathrm{Tr}(\hat{\rho}_i \hat{\Gamma}_j) P^{\Lambda}(i)}{\sum_{i,j} \mathrm{Tr}(\hat{\rho}_i \hat{\Gamma}_j) P^{\Lambda}(i)} \quad .$$

$$= \frac{\mathrm{Tr}(\hat{\rho}_i \hat{\Gamma}_j) P^{\Lambda}(i)}{\mathrm{Tr}(\hat{\rho}\hat{\Gamma})} \qquad \text{(A 5)}$$

where $\hat{\rho}$ is defined by (12). If we now introduce $\hat{\Lambda}_i$ by defining it as being proportional to $P^{\Lambda}(i)\hat{\rho}_i$, which is consistent with (10), and define $\hat{\Lambda}$ by (2), we find that (A 5) reduces to

$$P^{\Lambda\Gamma}(i,j) = \frac{\mathrm{Tr}(\hat{\Lambda}_i \hat{\Gamma}_j)}{\mathrm{Tr}(\hat{\Lambda}\hat{\Gamma})} \qquad \text{(A 6)}$$

in agreement with (1).